\documentclass[aps,physrev,print,superscriptaddress]{revtex4-2}

\usepackage{graphicx}
\usepackage{dcolumn}
\usepackage{bm}
\usepackage[hidelinks]{hyperref}
\usepackage[mathlines]{lineno}
\linenumbers\relax 
\usepackage{amssymb}
\usepackage{amsmath}
\nolinenumbers  
\usepackage{xcolor}
\usepackage[figure]{hypcap}
\usepackage[version=4]{mhchem}
\usepackage{subscript}
\usepackage{siunitx}
\DeclareSIUnit\Ions{ions}
\DeclareSIUnit\electronvolt{eV}
\DeclareSIUnit\electrons{e^{-}}
\DeclareSIUnit\grooves{grooves}
\usepackage{placeins}
\usepackage{chemmacros}
\chemsetup{modules = {reactions,isotopes} , greek   = upgreek ,formula = chemformula} 
\DeclareSIUnit\u{u}

\begin{document}

\preprint{APS/123-QED}

\title{\textbf{Laser fluence-dependent production of molecular thorium ions in different charge states for trapped-ion experiments}}

\author{Jonas Stricker}
\email[]{Contact author: jostrick@uni-mainz.de}
\homepage[]{superheavies.de}
\affiliation{Department of Chemistry - TRIGA Site, Johannes Gutenberg University Mainz, 55099 Mainz, Germany}
\affiliation{PRISMA$^+$ Cluster of Excellence, \\ Johannes Gutenberg University Mainz, 55099 Mainz, Germany}
\affiliation{Helmholtz Institute Mainz, 55099 Mainz, Germany}

\author{Jean Velten}
\affiliation{Department of Chemistry - TRIGA Site, Johannes Gutenberg University Mainz, 55099 Mainz, Germany}

\author{Valerii Andriushkov}
\affiliation{PRISMA$^+$ Cluster of Excellence, \\ Johannes Gutenberg University Mainz, 55099 Mainz, Germany}
\affiliation{Helmholtz Institute Mainz, 55099 Mainz, Germany}
\affiliation{QUANTUM, Johannes Gutenberg University Mainz, 55099 Mainz, Germany}

\author{Lennard M. Arndt}
\affiliation{Department of Chemistry - TRIGA Site, Johannes Gutenberg University Mainz, 55099 Mainz, Germany}

\author{Dmitry Budker}
\affiliation{PRISMA$^+$ Cluster of Excellence, \\ Johannes Gutenberg University Mainz, 55099 Mainz, Germany}
\affiliation{Helmholtz Institute Mainz, 55099 Mainz, Germany}
\affiliation{QUANTUM, Johannes Gutenberg University Mainz, 55099 Mainz, Germany}
\affiliation{University of California, Berkeley, California 94720, USA}

\author{Konstantin Gaul}
\affiliation{Helmholtz Institute Mainz, 55099 Mainz, Germany}
\affiliation{QUANTUM, Johannes Gutenberg University Mainz, 55099 Mainz, Germany}

\author{\\Dennis Renisch}
\affiliation{Department of Chemistry - TRIGA Site, Johannes Gutenberg University Mainz, 55099 Mainz, Germany}
\affiliation{PRISMA$^+$ Cluster of Excellence, \\ Johannes Gutenberg University Mainz, 55099 Mainz, Germany}
\affiliation{Helmholtz Institute Mainz, 55099 Mainz, Germany}

\author{Ferdinand Schmidt-Kaler}
\affiliation{PRISMA$^+$ Cluster of Excellence, \\ Johannes Gutenberg University Mainz, 55099 Mainz, Germany}
\affiliation{Helmholtz Institute Mainz, 55099 Mainz, Germany}
\affiliation{QUANTUM, Johannes Gutenberg University Mainz, 55099 Mainz, Germany}

\author{Azer Trimeche}
\affiliation{QUANTUM, Johannes Gutenberg University Mainz, 55099 Mainz, Germany}

\author{Lars von der Wense}
\affiliation{QUANTUM, Johannes Gutenberg University Mainz, 55099 Mainz, Germany}

\author{Christoph E. Düllmann}
\affiliation{Department of Chemistry - TRIGA Site, Johannes Gutenberg University Mainz, 55099 Mainz, Germany}
\affiliation{PRISMA$^+$ Cluster of Excellence, \\ Johannes Gutenberg University Mainz, 55099 Mainz, Germany}
\affiliation{Helmholtz Institute Mainz, 55099 Mainz, Germany}
\affiliation{GSI Helmholtzzentrum für Schwerionenforschung GmbH, 64291 Darmstadt, Germany}

\collaboration{TACTICa Collaboration}
\noaffiliation

\date{\today}

\begin{abstract}

Thorium ions and molecules, recognized for their distinctive nuclear and atomic attributes, are central to numerous trapped-ion experiments globally. Our study introduces an effective, compact source of thorium atomic and molecular ions produced via laser ablation of microgram-scale, salt-based samples. We thoroughly analyze the variety of ion species and charge states generated at varying laser fluences. Utilizing 10\,µg of thorium fluoride crystals and laser fluences between  $1.00 - 7.00$\,J$\cdot$cm$^{-2}$ we produce thorium molecular ions \ce{^{232}ThF_x^n+} (with $x= 0 - 3$ and charge states up to $n = 3$), including \ce{ThF^2+} and \ce{ThF^3+}. These species are particularly relevant for spectroscopy; \ce{ThF^3+} is valuable due to its stable closed-shell configuration, while \ce{ThF^2+}, which is isoelectronic to \ce{RaF}, offers a unique probe for studying nuclear structure and fundamental symmetries due to its simple electronic structure with a single unpaired electron. 
Density functional theory-based calculations of the electric charge distribution aid in understanding the formation of the observed species.  Positive charges are found to be mainly located on the thorium atom, which points to an inhibition of immediate Coulomb explosion, thus supporting the existence of the observed ion species. The simplicity of the method and similarities among the actinide elements suggest that our approach will also be applicable to other actinide species.

\begin{description}
\item[keywords] 
Laser ablation; charge stripping; mass spectrometry; thorium fluoride; actinide molecules in higher charge states.
\end{description}
\end{abstract}

\maketitle
\section{\label{sec:introduction}Introduction}
The element thorium, particularly the isotope \isotope{229,Th} is a central focus of recent research due to the exceptionally low-energy nuclear excited state $^{229m}$\ce{Th}, located above the ground state at 8.355729193198(8)\,eV  {\cite{Tiedau2024,Elwell2024,zhang2024frequency}}. The recent excitation from the nuclear ground state via VUV-laser light opens the door to the development of a nuclear clock {\cite{Peik2003,Rellergert2010,Campbell2012}} and the investigation of potential temporal variations of fundamental constants {\cite{Flambaum2006,Beeks2024fine}}. Such temporal variations of fundamental constants may arise from galactic ultralight bosonic dark matter or scalar dark matter \cite{Antypas2021} and can be investigated in thorium ions or molecular ions. In particular, thorium molecular ions exhibit significant sensitivity enhancements to various beyond-the-standard-model phenomena such as electric dipole moments which are a signature of charge-parity violation (CP-violation)\cite{meyer2008,skripnikov2013,fleig:2014,flambaum:2014,denis2015,skripnikov2015,denis:2016,Hutzler2020}.
There is a siginificant interest in thorium molecular ions with isotope $^{229}$Th, such as \ce{$^{229}$ThOH^+} and \ce{$^{229}$ThF^+} due to possibly large enhancements of CP-violation by nuclear structure via potential octupole deformations \cite{flambaum:2014,skripnikov2015,Flambaum2019}. 

While thorium monofluoride in the charge state of 1+ is already center of a variety of spectroscopy experiments \cite{Barker2012,Ng2022,Zhou2019}, the production of thorium monofluoride and molecular actinide ions in higher charge states has barely been studied.
Molecules like \ce{ThF^3+} appear promising in the use of spectroscopy experiments due to a stable closed-electronic-shell configuration\cite{zulch:2022}. Furthermore, \ce{ThF^2+} is isoelectronic to \ce{RaF} \cite{zulch:2022}. RaF has a simple electronic structure with essentially one unpaired electron above energetically well separated closed electronic shells, and therefore provides unique opportunities for the investigation of nuclear structure and fundamental symmetries \cite{isaev:2010,Isaev2013,Udrescu2021,wilkins:2023,GarciaRuiz2020}. \ce{ThF^2+} could combine the advantages of actinide molecules \cite{ArrowsmithKron2024} discussed above with advantages of \ce{RaF} and ions, which are comparatively easy to trap and cool.

Laser ablation from a target is one of the most commonly used offline ion-production methods for loading thorium ions into a Paul or Penning trap. Due to the significantly longer half-life of $^{232}$Th ($t_{1/2}=1.4\cdot10^{10}$ years) compared to, e.g., $^{229}$Th ($t_{1/2}=7.8\cdot10^{3}$ years), the specific activity of $^{232}$Th is comparatively small, making the handling of larger quantities of material easy. Therefore, typically, experiments of this kind start with the ablation of $^{232}$Th in different chemical forms (e.g., from metal targets). Laser ablation of $^{232}$Th ions was demonstrated in \cite{Campbell2009, Zimmermann2012,Troyan2013,Borisyuk2017,Piotrowski2020}, as well as in the Trapped And Cooled Thorium Ion spectroscopy via Calcium (TACTICa) project \cite{GrootBerning2018, Stopp2019}. A challenge of laser ablation of rare and short-lived isotopes such as $^{228,229,230}$Th is the very limited availability of material and the larger specific activity, making the handling of larger quantities of material practically impossible. For laser ablation of $^{229}$Th, Th-containing salt-based samples have been ablated directly \cite{Campbell2011,Thielking2018}. When producing atomic thorium ions, sufficiently high laser power can result in ions attaining charge states higher than 1+. Laser ablation for thorium ions in the 3+ charge state was reported by Campbell et al. \cite{Campbell2009, Campbell2011} and for molecular thorium oxide ions in the 2+ charge state (\ce{$^{232}$ThO^2+}) by Li et al. \cite{Li2024}. The laser power density plays a decisive role in the ablation process, as it not only influences the production of higher charge states, but also determines the formation of thorium oxide or pure thorium clusters when appropriately optimized \cite{Fischer2025}.

Here, we aim at the production of thorium molecular ions, and study in detail how to optimize the ablation parameters accordingly for desired production of molecular ion species in different charge states. One method is that pre-existing impurities in the thorium foil directly lead to the formation of desired molecules such as thorium oxide (\ce{ThO+}) \cite{Li2024,fischer2024}. Also, in-flight chemical reactions can be employed, where atomic ions produced during laser ablation interact with a buffer gas \cite{Nguyen2019}. We optimize the laser pulse fluence and directly use target materials that already contain the desired ions, such as using \ce{ThF4} crystals to generate \ce{ThF+} ions. As a result, we show, that this mitigates the requirement of buffer-gas mixtures containing carbon tetrafluoride (\ce{CF4}) which are introduced during the ablation process to produce singly charged thorium monofluoride (\ce{ThF+}) ions \cite{Nguyen2019,Au2023,Au2024} or larger molecules like \ce{ThF_{2-4}+} \cite{Bubas2022}, thus maintaining UHV conditions during the entire production process. Another asset is that our method is not limited to singly-charged molecules \cite{Wanless1971,Hutzler2012}, due to electron capture in the buffer gas reactions, but we demonstrate the creation of higher charge states. Also, our method is favorable, to produce multiply-charged ions ($n+ > 2$) experimentally, as compared to a much more demanding production process of, e. g., triply charged actinide molecules by ion - ion collisions \cite{Schroeder1999}. 

In this work, we present a novel and practical pulsed laser ablation and ionization source as a table-top device for producing actinide molecular ions in charge states up to 3+ as part of the TACTICa project at Johannes Gutenberg University Mainz, Germany \cite{Haas2019}. By varying the laser fluence (\textit{$\Psi$}), we selectively produce any of several thorium fluoride molecules in different charge states. Quantum chemical calculations of charge distributions in \ce{ThF_x^{n+}} ions are performed to support understanding of the feasibility of producing multiply-charged molecular thorium fluoride ions. The direct production of exotic molecular ions through the ablation of salt-based samples enables experiments in an ultra-high vacuum environment without the need for differential pumping or reliance on gas-phase reactions, targeting for high-precision measurements. 

\section{Methodology}
\subsection{Experimental setup and calibration of the time-of-flight mass spectrometer}

\begin{figure*}[ht]
\includegraphics[width=1\linewidth]{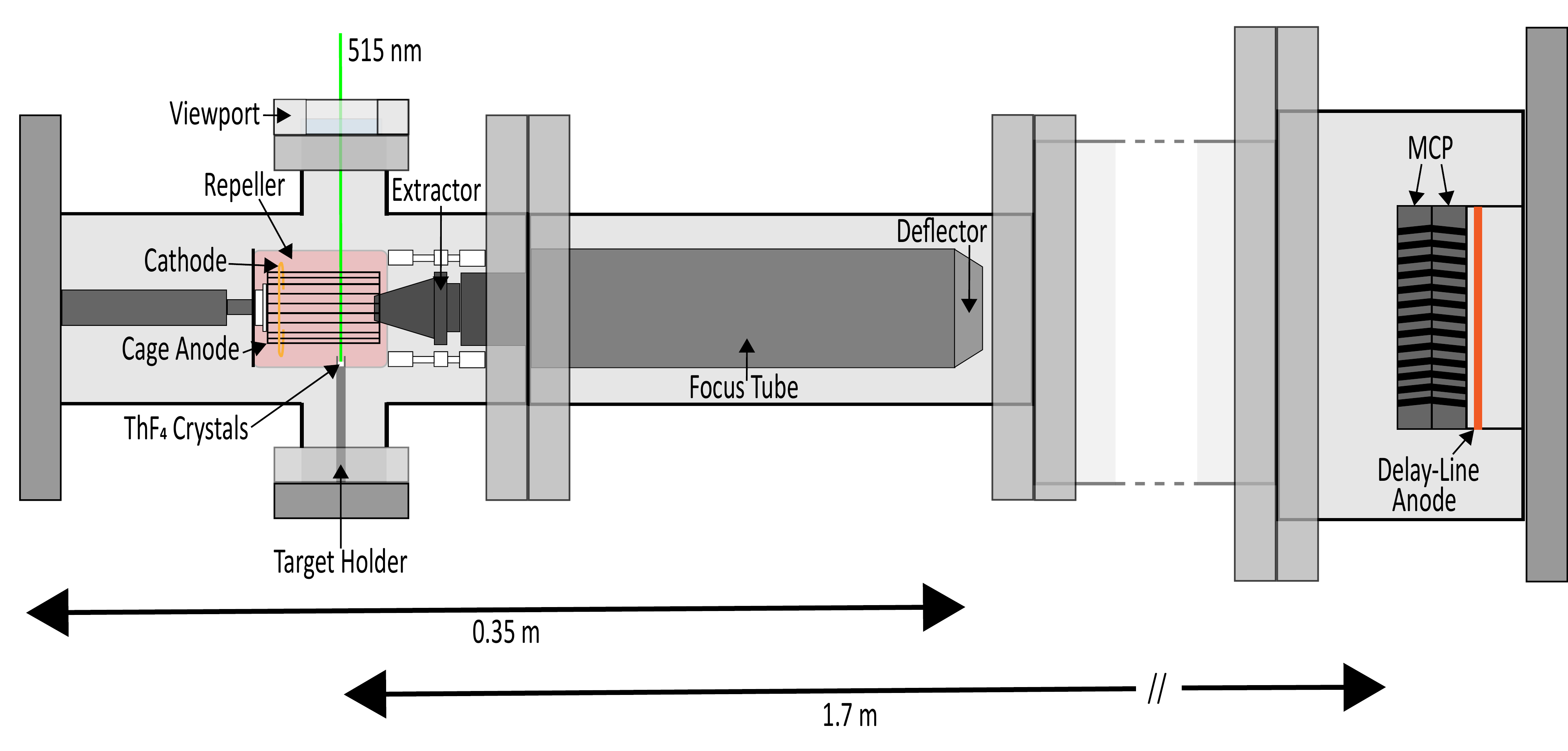}
\caption{\label{fig:1}Schematic of the ToF mass spectrometer, including the modified ion gun (\textit{SPECS IQE 12/38}), used for salt based and metallic targets, and two MCPs in combination with a delay-line anode (\textit{Reontdek DLD40f}) for ion detecting (not to scale). The modular setup would allows an easy installation of an ion trap between the ion gun and detector.}
\end{figure*}

We have built a new time of flight mass spectrometer to investigate microgram targets ablated by laser light with different fluences to produce molecular actinide-ion species for trapped-ion experiments. Laser ablation is employed to vaporize or directly ionize material from the target material within an ultra-high vacuum environment, inside of a modified ion sputter gun (\textit{SPECS IQE 12/38}). The ion gun consists of a filament/cathode, for emitting electrons for electron impact ionization, a repeller, to repel the produced electrons, a cage anode, to set the starting kinetic energy of the produced ions, an extractor for extraction the ions and ion optics, a focus tube and two electrostatic deflectors, for manipulating the resulting ion beam (see FIG. \ref{fig:1}). After traversing a drift tube, ions are detected by two multi-channel-plates (MCP) in combination with a delay-line anode (\textit{Reontdek DLD40f}). 
The MCP is operated with a high gain ($>10^7$), and the delay-line anode system includes a carefully tuned drift region with an optimized potential difference of +300\,V between the back end of the MCP and the anode wires. Therefore, reliable signal formation of heavy ion species, such as \ce{ThF3^n+}, is ensured under these conditions.
The flight path of the ions is approximately 1.7\,m long. For all experiments and for the calibration of the apparatus, the same experimental parameters of the ion gun and the MCPs were used, which can be found in TABLE \ref{tab:Gun}.

\begin{table}[h]
\caption{\label{tab:Gun}
The table shows the applied voltages (U in kV) of the ion gun and MCP detector.}\begin{ruledtabular}
\begin{tabular}{cccccc|cc}
Cage anode & Extractor & Focus tube  &  & Filament & & MCP & \\
energy & extraction & focus 1 & focus 2 & emission & & front end & back end \\
\hline
 1.50 &  1.16 & 1.23 & 0 & 10 mA & &  - 2.40 & 0.30 \\
\end{tabular}
\end{ruledtabular}
\end{table}

For the ablation of metallic and salt based targets from a ceramic target holder the ion gun was modified, see FIG.\,\ref{fig:1}. A \textit{Coherent FLARE NX71 515-0.6-2} laser, operating at a wavelength of $\lambda$\,=\,515\,$\pm$\,5\,nm with a pulse duration of 1.3\,$\pm$\,0.2\,ns and an energy output of $E$\,=\,300\,$\pm$\,15\,µJ, employed for laser ablation. The laser's focused beam, with a diameter of 70$\pm$7\,µm, vaporizes the target material, producing a plasma consisting of ions, neutrals, and electrons.

To produce ions over a large range in mass-to-charge ratio for calibration purposes we ablated different non-actinide samples. For the production of non-actinide ions and operating at non-ionizing laser fluences electron impact is used for ionization. At non-ionizing laser fluences laser ablation vaporizes the target material and produces neutral species, which are moving towards the cage anode. The neutral species are than ionized by electron impact, stemming from the electron-emitting iridium-coated tungsten cathode, inside of the cage anode. The emitted electrons are accelerated to the tungsten cage anode and then repelled by the repeller so that the highest electron density is in the center of the repeller (see FIG.\,\ref{fig:1}). Following ablation and ionization, ions are extracted by the extractor at a voltage of 1.5\,kV, for best focus of the ion beam onto the detector. The resulting ion beam is focused into a diameter of 160 µm using the two electrostatic lenses of the focus tube. The ion flight path can be manipulated by two electrostatic deflectors, which was not used in the experiments.

\begin{figure*}[h]
\includegraphics[width=1\linewidth]{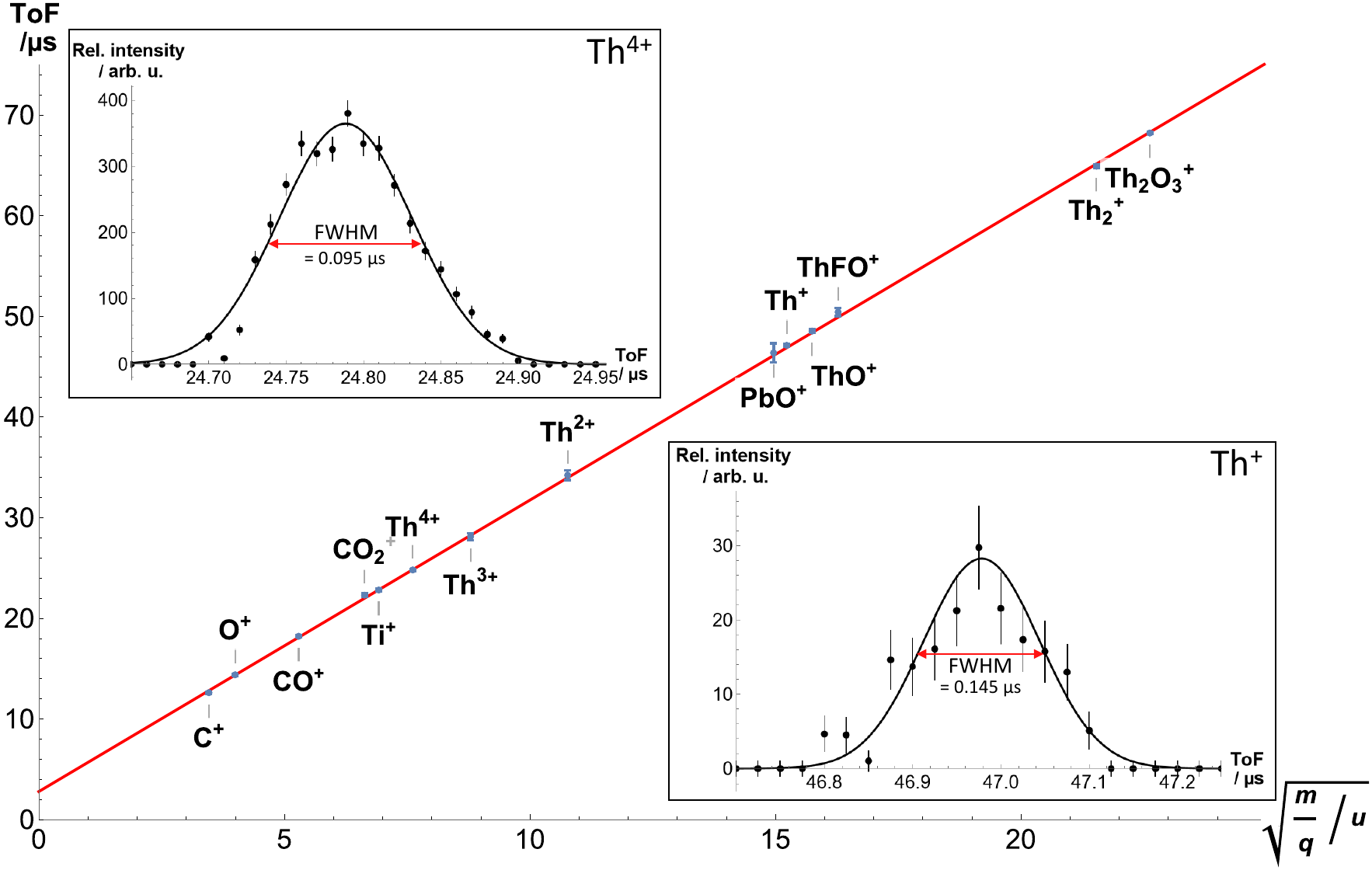}
\caption{\label{fig:X}{Calibration fit of different compounds relating the time of flight (ToF) with $\sqrt{m/q}$. The insets show exemplarily the ToF spectra of  \ce{Th^4+} and \ce{Th^+}.}}
\end{figure*}

All measurements for non-actinide species were performed with the electron-emitting cathode on. By measuring and analyzing ToF-spectra of non-actinide ions, a calibration was be obtained (see FIG.\,\ref{fig:X}). Actinide-ion species were assigned according to their mass-to-charge ratio and thus identified. To calibrate the setup, the proportional dependence between flight time and mass-to-charge ratio is used (see Eq. \ref{Fitgleichung}). This results in the following fit equation, in which $a$ is a proportionality factor and $b$ is a time offset:

\begin{equation}
    \mathrm{ToF}=a \cdot \sqrt{\frac{m}{q}}+b.
\label{Fitgleichung}
\end{equation}
\FloatBarrier

Samples of metallic $^{232}$\ce{Th}, lead and titanium were used to set calibration points. In addition, carbon, oxygen and molecular combinations of the two elements were also found in all metallic samples. These points also served as calibration aids for the different spectra.
A weighted fit is performed, with the weight of a data point corresponding to the reciprocal quadratic value of the ToF uncertainty. Plotting ToF against the $\frac{m}{q}$ ratio results in the fit curve shown in FIG.\,\ref{fig:X}.

The resulting fit parameters are $a = 2.892(6)\,\frac{\upmu \mathrm{s}}{\sqrt{\mathrm{u}}}$ and $b = 2.72(7)\,µs$, where b represents the timing offset of the arriving ions. In the following, time-of-flight spectra are converted into mass-over-charge spectra, with the charge being $q = 1$,  with the aid of the determined fit equation as follows:

\begin{equation}
    \frac{m}{q}=\left( \frac{\mathrm{ToF}-b}{a} \right)^2.
\label{umrechnung}
\end{equation}
\FloatBarrier

The mass resolving power $R$ of our time-of-flight spectrometer is obtained as
\begin{equation}
R = \frac{m}{\Delta{m}}\approx \frac{\mathrm{ToF}}{2\ \Delta\mathrm{ToF} }\approx150,
\label{resolv}
\end{equation}
\FloatBarrier

where $\Delta\mathrm{ToF}$ is defined as the FWHM of the peaks in the ToF-spectrum and experimentally determined. The second equation is a result of Gaussian error propagation, considering that the mass resolving power is dominated by the widths of the peaks in the ToF-spectrum. Experimentally we find that the mass resolving power also depends on the laser fluence, which was kept at its maximum for FIG.\,\ref{fig:X}. Varying the fluence allowed us obtaining high enough resolution for an unambiguous identification of each species shown in FIG. \ref{fig:2}.

For the ablation of thorium fluoride, the target material, consisting of 10\,µg of \ce{\isotope{232,Th}F4} crystals, is directly synthesized  directly synthesized in a notch of a MARCOR ceramic target rod by the reaction of thorium nitrate and hydrofluoric acid. Detailed information about the synthesis and the purity of \ce{ThF4} can be found in the Supplemental Material \cite{SM}. Given the smallness of the sample, the approach is directly applicable for the use of exotic thorium isotopes like \isotope{229,Th} or other rare actinide and non-actinide samples. After ablation in combination of direct ionization by the laser the produced molecular or atomic ions are accelerated by the cage anode and extracted by the extractor by a voltage of 1.5\,kV, focused and detected by the detector. 

For producing ions from the \ce{ThF4} target the electron beam was activated. However, the deactivation of the electron beam did not influence the ToF results for actinide ion species. Apparently, the ionization process of Th ions and molecular ions is fully dominated by the laser pulse induced processes.

\subsection{Quantum chemical calculations of charge distributions in \ce{ThF_x^n+} molecular ions}
From previous studies of triply charged uranium monofluoride (\ce{UF^3+})\cite{Schroeder1999} we expect \ce{ThF^n+} to be stable for $n\leq3$. Moreover, the thermodynamic stability of similar actinide monofluoride ion species was theoretically studied before \cite{zulch:2022,zulch:2023}. Based on these studies, we assume that additional fluoride atoms carry essentially no positive charge. We expect that \ce{ThF_x^n+} with $x+n\leq5$ may be (meta-)stable and, therefore, could be observed in the mass spectrum. To support this assumption we computed the charge distribution in the produced molecules. Accurate calculations of the thermodynamic stability of the studied molecules and investigations of their spectroscopic suitability for studies of fundamental physics will be subject of a separate study \cite{zulch:2025}.

All DFT  calculations were performed with a modified version
\cite{gaul:2020,zulch:2022} of a two-component program \cite{wullen:2010} based
on Turbomole \cite{ahlrichs:1989}. We employed the exchange correlation
functional by Perdew, Burke and Ernzerhof (PBE) \cite{perdew:1996} in a hybrid
version with 50\,\% Fock exchange (PBE50) \cite{bernard:2012}. All calculations
were performed within a quasi-relativistic complex generalized Kohn-Sham (cGKS)
framework including relativistic effects on the two component zeroth order regular approximation
(2c-ZORA) level. 2c-ZORA was employed with a damped model potential to alleviate the gauge dependence
\cite{wullen:1998,liu:2002}. The Hilbert space was sampled with atom-centered
Gaussian basis functions using the core-valence basis set of triple-$\zeta$
quality by Dyall (dyall.cv3z) \cite{dyall:2002, dyall:2006}. The nuclear charge density distribution was
modeled as a normalized spherical Gaussian 
$\varrho_K \left( \vec{r} \right) = \frac{\zeta_K^{3/2}}{\pi ^{3/2}}
\text{e}^{-\zeta_K \left| \vec{r} - \vec{r}_K \right| ^2}$ with $\zeta_K =
\frac{3}{2 r^2 _{\text{nuc},K}}$. The root-mean-square radius
$r_{\text{nuc},K}$ was chosen as suggested by Visscher and Dyall employing the 
isotopes $^{19}$F and $^{232}$Th \cite{visscher:1997}. Electronic densities were
converged until the change of the total energy between two consecutive cycles in
the self-consistent field procedure was below $10^{-10}\,E_\mathrm{h}$.
Molecular structures were optimized until the change of the norm of the 
gradient with respect to nuclear displacements was below
$10^{-3}\,E_\mathrm{h}/a_0$ and the change of the total energy was below
$10^{-6}\,E_\mathrm{h}$. Obtained molecular structure parameters are provided
in the Supplemental Material \cite{SM}.
The supposedly lowest lying electronic states were found using the 
closed-shell species \ce{ThF^3+}, \ce{ThF2^2+} and \ce{ThF3+} as starting points
and adding or removing sequentially electrons. This method does not guarantee to find the global minimum, and therefore energetically lower lying electronic states may exist on the level of 2c-ZORA-PBE50/dyall.cv3z. However, we do not expect an essential change in the charge distribution in potential energetically lower-lying states. For reproducibility the obtained electronic states were
characterized by computing the reduced total electronic angular momentum
$\vec{J}_\mathrm{e}=\vec{L}+\vec{S}$, where $\vec{L}$, $\vec{S}$ are the electronic orbital and spin angular momenta respectively. In addition we computed the expectation value of $\hat{S}^2$, $S(S+1)$. Although $L_z$, $S_z$ and $S$ are not good quantum numbers in a
relativistic framework, they can be used to estimate the composition of the 
spin symmetry-broken determinant from configuration state functions of non-relativistic
symmetry (see also Refs. \cite{zulch:2022,zulch:2023}). Angular momenta of the 
computed states of \ce{ThF_x^n+} molecules are listed in the Supplemental Material \cite{SM}. 
For non-linear molecules the direction of $\vec{J}_\mathrm{e}$ is arbitrary and we report only the length of the angular momentum vector. Linear molecules were oriented with the molecular axis being aligned to the $z$-axis. The total angular momentum of diatomic molecules was aligned essentially to the $z$-axis as well. Therefore, for diatomic molecules  $\Omega=\left|\vec{J}_\mathrm{e}\right|$ and analogue for other angular momenta. For the optimized electronic states we computed the distribution of electrons over the nuclei using a simple Mulliken population analysis.

\begin{table}
\caption{Mulliken partial charges of \ce{ThF_x^n+} with $x,n=1,2,3$ molecular ions computed at the level of 2c-ZORA-PBE50/dyall.cv3z for the optimized electronic states (see Sec. II.B and Supplemental Material for details\cite{SM}).}
\label{tab:partialcharges}
\begin{ruledtabular}
\begin{tabular}{lSSSS}
Molecule & {$\delta_\mathrm{Th}/e$} & {$\delta_\mathrm{F1}/e$} & {$\delta_\mathrm{F2}/e$}  & {$\delta_\mathrm{F2}/e$}  \\
\hline
\ce{ThF+}     & +1.31 & -0.31& {-}  &{-}  \\
\ce{ThF^2+}  & +2.21 & -0.21& {-}  &{-}  \\
\ce{ThF^3+}  & +3.00 &  0.00& {-}  &{-}  \\
\ce{ThF2+}    & +1.64 & -0.32&-0.32 &{-}  \\
\ce{ThF2^2+} & +2.37 & -0.19&-0.19 &{-}  \\
\ce{ThF2^3+}& +2.79 & +0.23&-0.02 &{-}  \\
\ce{ThF3+}   & +2.02 & -0.34&-0.34 &-0.34\\
\ce{ThF3^2+} & +2.28 & +0.10&-0.19 &-0.19\\
\ce{ThF3^3+}& +2.59 & +0.23&+0.23 &-0.04\\
\end{tabular}
\end{ruledtabular}
\end{table}

\section{\label{sec:results}Results and discussion}

\begin{figure*}[ht]
\includegraphics[width=1\linewidth]{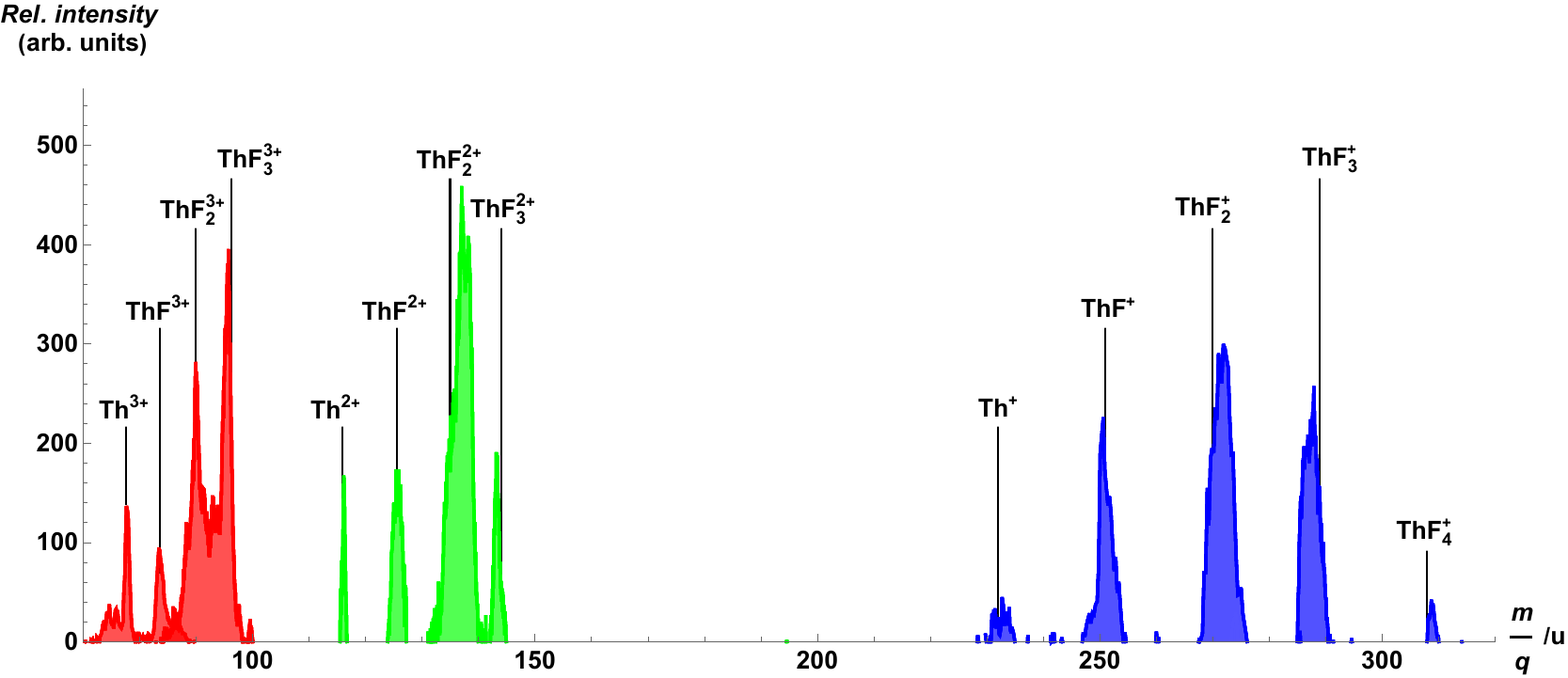}
\caption{\label{fig:2}Spectra of measured atomic and molecular Th ions in charge states 1+ (blue), 2+ (green), and 3+ (red). Nine individual spectra recorded at the fluences listed in TABLE \ref{tab:table1} are summed.}
\end{figure*}

The experiments revealed the formation of various charge species of thorium fluoride from laser ablation. As depicted in FIG.\,\ref{fig:2}, we observed that the mass-over-charge ratios of these ions suggest the presence of both atomic and molecular thorium ions as \ce{\isotope{232,Th}F_x^n+} with $x = 0 - 3$ and ionization states up to $n = 3+$ (see TABLE\,\ref{tab:table1}. A general trend is seen, that tri- and tetratomic molecular ions have a higher relative intensity compared to mono- and diatomic ions, with the highest relative intensity at \ce{ThF2^2+}. Through ablation of the thorium fluoride target by varying the laser power in combination with different laser spot sizes, it was possible to determine the optimum fluence ranges. The optimum fluence range in which the respective ion species achieves the highest intensity relative to all the observed species can be found in TABLE \ref{tab:table1}.

\begin{table}[ht] 
\caption{Production of atomic and fluoride-containing Th ions in charge states 1+ to 3+. The table shows the fluence ( $\Psi$ in J$\cdot$cm$^{-2}$) ranges in which the respective species dominates.}
\label{tab:table1}
\begin{ruledtabular}
\begin{tabular}{c|cccc}
\,\,\,\,\,\,\,\,\,\,\,\,Ion species\,\,\,\,\,\,\,\,\,\,\,\, & 3+ & 2+ & 1+& \\
\hline 
\ce{Th} &$6.75\pm0.35$ &  $4.00\pm0.20$ & $2.80\pm0.15$ &\\ 
\ce{ThF} &$5.20\pm0.30$ &  $2.40\pm0.25$ & $1.40\pm0.05$ &\\ 
\ce{ThF2} &$1.60\pm0.05$ &  $1.50\pm0.5$  &  $1.40\pm0.05$ &\\ 
\ce{ThF3} &$1.60\pm0.05$  &  $1.50\pm0.5$  &  $1.00\pm0.05$ &\\ 
\ce{ThF4} &- & - & $2.80\pm0.15$  \\
\end{tabular}
\end{ruledtabular}
\end{table}

In FIG. \ref{fig:2}, the gray vertical lines show the calculated m/q positions of the respective ions based on the calibration curve. The experimental peaks were identified by their centers. These lie  on the calculated masses within the experimental uncertainty. Minor deviations between the expected and observed peak positions may arise from detector saturation effects \cite{Mathew2022, Wei2020}, particularly for intense and closely-spaced peaks.  However, the purity of the crystals was confirmed by additional analysis, which excludes foreign ions (see \cite{SM}).

We conjecture that the ability to produce thorium molecular ions in higher charge states is primarily attributable to a) the fact that the initial three ionization energies of the thorium atom (6.30\,eV, 12.10\,eV, 18.32\,eV \cite{Kramida1999}) are lower than the dissociation energy of actinide fluoride bonds \cite{Schröder1999, zulch:2022, Franzreb2004}, and b) the charge being predominantly localized on the central thorium atom, which inhibits a Coulomb explosion. Both factors combined allow the central thorium atom to be ionized before dissociation of the \ce{Th-F} bond occurs. To support assumption b) we computed Mulliken partial charges of \ce{ThF_x^{n+}} ions as detailed in the computational methods section. The results are shown in Table \ref{tab:partialcharges}. These partial charges suggest that the charge of \ce{ThF_x^{n+}}, with $x+n<5$ is almost exclusively located at Th with the fluorine atoms being neutral or bearing slight negative charges. The calculations indicate that in \ce{ThF2^{3+}}, \ce{ThF3^{2+}}, and \ce{ThF3^{3+}} a small fraction of the charge may be moved to one of the fluorine atoms. This could be explained by the removal of an electron from a \ce{Th-F} bond which is in line with elongated \ce{Th-F} bonds for these species (see TABLE Supplemental Material \cite{SM}). However, the charge on the fluorine atoms seems to be small enough to inhibit immediate Coulomb explosion of these molecular species. A quantitative discussion of the thermodynamic stability of the observed species requires calculating energies of all possible dissociation channels or potential energy curves for all \ce{ThF_x^{n+}} which will be provided in a separate work \cite{zulch:2025}.

In Figure \ref{fig:2}, we examine the m$\cdot$q$^{-1}$ of $\approx$77, where both \ce{Th^3+} and \ce{ThF4^4+} could be identified as the produced species. Compared to observations of the formation of \ce{ThF^2+} at a fluence of 2.40\,J$\cdot$cm$^{-2}$, \ce{Th^3+} is the formed as a result of the complete dissociation of all bonds. This is underlined within the fluence ranges of 2.80 to 4.00\,J$\cdot$cm$^{-2}$ resulting in \ce{Th^+} and \ce{Th^2+} (see TABLE \ref{tab:table1}), while no \ce{ThF4^2+} could be observed at a fluence of 2.40\,J$\cdot$cm$^{-2}$. 
The trend of bond dissociation with laser fluence ranging from 2.00 to 6.75\,J$\cdot$cm$^{-2}$ is visible more likely to occur before leading to higher charge stages. In contrast, the process of bond dissociation is less dominant in the range of 1.00 to 2.00\,J$\cdot$cm$^{-2}$, and reaching molecular ion in higher charge states does not necessitate significant increases in fluence unlike for atomic thorium ions. This is particularly relevant as the thorium fluoride sample does not exhibit a perfect crystal structure, caused by the chemical synthesis reaction chose, resulting in the production of \ce{ThF_X^n+} instead of \ce{ThF4^n+} (both $ n = 1 - 3$) after ablation. A crystallographically grown thorium fluoride crystal will probably have a smaller defect density.
A direct statement about the dissociation process by analyzing the peak shape cannot be made, due to the extraction of the ions. Although dissociation processes could, in principle, lead to kinetic energy release and peak broadening, the use of lateral extraction at 1.5\,kV significantly suppresses energy-dependent ToF spread for energies as they emerge from dissociation processes. All observed peaks exhibit Gaussian shapes without discernible asymmetries.

The formation of atomic thorium (\ce{Th^n+}) and thorium monofluoride (\ce{ThF^n+}) ions is significantly influenced by the application of a laser fluence above 2.00\,J$\cdot$cm$^{-2}$. The relative intensities of the signals are comparably low for tri- and tetratomic ion species. During the ion production, a substantial ion loss occurs due to the ablation process itself at fluences above 2.00\,J$\cdot$cm$^{-2}$, which leads to ion absorption by the walls of the sample holder or other parts of the apparatus leading to low relative intensities. In contrast, at fluences below 2.00\,J$\cdot$cm$^{-2}$, the ablated ions are concentrated in a smaller plasma volume, leading to higher intensities.

The range of 2.00 to 3.00\,J$\cdot$cm$^{-2}$ represents the threshold between the processes of direct ionization and bond dissociation. This is demonstrated by the simultaneous laser-induced production of \ce{Th^+} and \ce{ThF4^+}. In this scenario, the majority of the energy goes into bond dissociation, while the residual energy facilitates production of \ce{ThF4^+}. These observations highlight the complex interplay between direct ionization, vaporization, and bond dissociation in the production of various ionic species. This indicates that different ionic species require distinct conditions for effective ionization, potentially affecting the overall efficiency and selectivity of the ion generation process.

The MCP detection efficiency depends on the ion impact velocity, which scales as $v\,\propto\,\sqrt{q/m}$. Heavier ions with lower charge states (e.g., \ce{ThF_3+}) have lower velocities than lighter or more highly charged species (e.g., \ce{ThF3^{3+}}), which results in a reduced detection efficiency. Experimental studies have shown that MCP detection efficiency can vary by up to a factor of two for ions whose impact velocities differ by a factor of $\sqrt{3}$~\cite{LadislasWiza1979, Brehm1995, Fraser2002}. While this may influence the absolute intensity ratios between ions of different charge states, it does not affect species identification or the interpretation of relative production trends within a given charge state.

Further investigation is needed to determine if production of higher charge states as 3+ in molecular thorium ions is possible or, if Coulomb repulsion precludes the formation of such ions.

 \section{\label{sec:Conclusion}Outlook for further work}

Direct laser ablation of thorium fluoride crystals effectively generates atomic thorium ions and thorium fluoride ions in charge states up to 3+. This newly designed tabletop ToF source can be used to load atomic and molecular ions of thorium with chosen charge states into ion-traps due to its modularity (see FIG. \ref{fig:1}). 

These experiments highlight the potential of microgram thorium fluoride crystals as laser ablation targets for applications in the exploration of physics beyond the standard model (BSM). While \isotope{232,Th} was used in this work due to its availability and long half-life, this method can be adapted to more rare isotopes such as \isotope{228,Th}, \isotope{229,Th}, and \isotope{230,Th}, through microgram scale synthesis techniques.

This approach can likely be extended to the production other actinide molecules \ce{AnF_x^n+}, where the dissociation energy of the actinide fluoride bond is higher than the ionization energy of the actinide atom and also, where the charge would similarly reside on the central actinide atom \cite{zulch:2022}. Several molecules of these type are also relevant to BSM physics \cite{ArrowsmithKron2024}, highlighting the potential for broader applicability and paving the way for future studies and technological advancements in this area.

\section{Acknowledgements}
This work has been supported by the Cluster of Excellence “Prediction Physics, Fundamental Interactions, and Structure of Matter$^+$” (PRISMA$^+$ + ExNet-020). Funded by the German Research Foundation (DFG) under TACTICa (Project Nr. 495729045). K.G. thanks the Fonds der Chemischen Industrie (FCI) for generous funding through a Liebig fellowship. We gratefully acknowledge computing time at the supercomputer MOGON 2 at Johannes Gutenberg University Mainz (hpc.uni-mainz.de), which is a member of the AHRP (Alliance for High Performance Computing in Rhineland Palatinate,  www.ahrp.info) and the Gauss Alliance e.V.
Fruitful discussions with Lutz Schweikhard, Paul Fischer and Tom Kieck are acknowledged.

\section{Author contributions}
J.S.: conceptualization, formal analysis, investigation, writing - original draft. 
J.V.: formal analysis, investigation, writing - review \& editing.
V.A.: writing - original draft.
L.M.A.: formal analysis, investigation.
D.B.: resources, funding acquisition, supervision, writing - review \& editing. 
C.E.D.: resources, funding acquisition, supervision, writing- review \& editing. 
K.G.: resources (theoretical), funding acquisition, investigation (theoretical), writing- original draft. 
D.R.: resources, writing - review \& editing. 
F.S.-K.: funding acquisition, supervision, writing - review \& editing.
A.T.: investigation, writing - review \& editing.
L.vdW.: resources, writing - review \& editing.

\bibliography{References}

\end{document}